\documentclass[pra,twocolumn,graphicx]{revtex4}
\usepackage{epsfig,amsmath}

\begin{document}

\title{High-order harmonic generation with a strong laser field and an
attosecond-pulse train: the Dirac Delta comb and monochromatic limits}
\author{C. Figueira de Morisson Faria$^{1}$ and P. Sali\`eres$^2$}
\affiliation{$^1$Centre for Mathematical Science, City
University,
Northampton Square, London EC1V 0HB, United Kingdom\\
$^2$CEA-SPAM, B\^{a}t. 522, Centre d'Etudes de Saclay, F-91191
Gif-Sur-Yvette, France}

\date{\today}

\begin{abstract}
In recent publications, it has been shown that high-order harmonic
generation can be manipulated by employing a time-delayed attosecond
pulse train superposed to a strong, near-infrared laser field. It is
an open question, however, which is the most adequate way to
approximate the attosecond pulse train in a semi-analytic framework.
Employing the Strong-Field Approximation and saddle-point methods,
we make a detailed assessment of the spectra obtained by modeling
the attosecond pulse train by either a monochromatic wave or a
Dirac-Delta comb. These are the two extreme limits of a real train,
which is composed by a finite set of harmonics. Specifically, in the
monochromatic limit, we find the downhill and uphill sets of orbits
reported in the literature, and analyze their influence on the
high-harmonic spectra. We show that, in principle, the downhill
trajectories lead to stronger harmonics, and pronounced enhancements
in the low-plateau region. These features are analyzed in terms of
quantum interference effects between pairs of quantum orbits, and
compared to those obtained in the Dirac-Delta limit.
\end{abstract}

\maketitle

\section{Introduction}

\label{intro}

One attosecond (10$^{-18}s$) is roughly the time it takes for light
to travel through atomic distances. This fact makes high-frequency
pulses of attosecond duration a very powerful tool for resolving or
even controlling dynamic processes occurring at the atomic scale
\cite{AdiM2004}. Indeed, in the past few years, such pulses have
caused a breakthrough in metrology, with applications as diverse as
resolving the motion of bound electrons \cite {atto1}, exciting
inner-shell electrons \cite{atto2}, inducing vibrational wavepackets
in molecules \cite{atto3}, or controlling electron emission
\cite{atto4}.

Attosecond pulses have been predicted theoretically in the mid-1990s
\cite {saclay96} and have become experimentally feasible a few years
later \cite {attoexptrain,attoexp,mairesse}. They are a direct
consequence of the fact that high-order harmonics, generated by the
interaction of a strong laser field (intensities of the order of
$10^{14}\mathrm{W/cm}^{2}$) with a gaseous sample, are almost phase
locked \cite{saclay96}. Hence, by superposing sets of high
harmonics, it is possible to produce attosecond pulses.
Specifically, there exist two main scenarios: either one obtains an
attosecond-pulse train, by employing long input pulses
\cite{attoexptrain,mairesse}, or isolated attosecond pulses, from
few-cycle driving fields \cite{attoexp,fewcyclerev,atto1}. Both
situations have been widely exploited in the literature, and
sometimes the attosecond pulses appear in combination with
additional driving fields.

For instance, recently, an attosecond-pulse train superposed to a strong,
infrared, linearly polarized field has been employed to control
high-harmonic generation (HHG) \cite{atthhg2004,dpg} and above-threshold
ionization (ATI) \cite{attati2005}. Such a train exhibited a time-delay $%
t_{d}$ with respect to the infra-red field. By varying this delay,
one could influence several features, such as the intensities,
resolutions, and maximal energies in the high-harmonic or
photoelectron spectra. This scheme has been proposed theoretically
\cite{atthhg2004,attati2005} and realized experimentally
\cite{attati2005,dpg}, and the key idea behind it is to provide an
additional pathway, which can be controlled, for an electronic wave
packet to be released in the continuum.

This can be understood in view of the physical mechanisms behind
both phenomena. At a time $t^{\prime },$ an electron is freed by
tunneling or multiphoton ionization.$\;$Subsequently, it propagates
in the continuum, gaining energy from the field, and it is driven
back towards its parent ion, to which it returns at a time $t$. If
the electron recombines or rescatters, there is either emission of
high-frequency radiation (i.e., HHG), or of high-energy
photoelectrons (i.e., high-order ATI) \cite{tstep}. For low-order
ATI peaks, the electron reaches the detector without rescattering.
An attosecond pulse train allows the electron to reach the continuum
by absorbing high-energy photons and being able to overcome the
ionization potential. Additionally, since they are of very short
duration, the attosecond pulses provide a tool to control the
instant at which the electron is being ejected. This has direct
consequences in the momentum it acquires from the field when it is
released \cite{attati2005}, and also a strong influence on the
kinetic energy the electron has upon return \cite{atthhg2004,dpg}.
Therefore, it also affects the spectra.

The results in \cite{atthhg2004,dpg} and \cite{attati2005} have been
obtained from the numerical solution of the time-dependent
Schr\"{o}dinger equation, and the attosecond-pulse train has been
modeled by the superposition of five harmonics ($\Omega
_{11}=11\omega $ to $\Omega _{19}=19\omega $). These are realistic
assumptions, and well within the available experimental data.
However, in order to interpret the results obtained, the
attosecond-pulse train has been approximated by a monochromatic wave
of frequency $\Omega _{15}=15\omega ,$ which is the central
frequency of the group of harmonics in question, and a classical
model for an electron ejected by such a wave and propagating in a
strong, low-frequency field was constructed. These simplifications
have enormously facilitated the understanding of the ionization
mechanisms, and, specifically, have shown how to manipulate
ionization by varying the time delay $t_{d}.$

In a previous publication, we have also investigated the ATI and HHG
spectra from an atom irradiated by a time-delayed attosecond-pulse
train superposed to a near-infrared laser field \cite{FSVL2006}.
 Specifically, we have considered the transition amplitudes for both
phenomena within the Strong-Field Approximation, and took the
attosecond pulse train to be a sum of Dirac-Delta functions in the
time domain. These assumptions have allowed an almost entirely
analytical treatment, which has been vital for a clear understanding
of the problem. Indeed, we were able to investigate the influence of
the attosecond pulses on the spectra in far more detail than in
\cite{atthhg2004,dpg,attati2005}, and to interpret the results
obtained in terms of quantum interference effects.

To a very large extent, our results, as well as their physical
interpretation, agree with those in
\cite{atthhg2004,dpg,attati2005}. In fact, we have obtained, for
very specific time delays, enhancements of more than one order of
magnitude in the harmonic spectra similar to those reported in
\cite{atthhg2004}. Furthermore, as the time delay increases, the
enhancements move gradually from the high-order harmonics towards
lower frequencies, until they only affect the low-plateau region.
Finally, for ATI, we have observed an identical behavior of the
yield with the time delay as in \cite{attati2005}, namely that it
extends towards higher energies if $\omega t_d=n\pi$, and that it
decays most quickly, as the frequency increases, for $\omega
t_d=(2n+1)\pi/2$.

There exist, however, a few discrepancies, especially for HHG,
between our results and those in \cite{atthhg2004}. First, the
above-mentioned features occur for different phases $\phi=\omega
t_d$. Second, the explanations for such features are slightly
different. In \cite{FSVL2006}, we related the enhancements in the
spectra to a particular pair of orbits, for which the excursion
times $\Delta t=t-t^{\prime}$ of the electron in the continuum were
very short. Therefore, the spreading of the electronic wave packet
in the continuum would be very reduced. Hence, the overlap of this
wave packet, upon return, with the bound state it left, would be
considerable, leading to very prominent harmonics. We have shown
that the maximal kinetic energy the electron may have upon return,
for this pair of orbits, was very much dependent on the delay
between the infra-red field and attosecond-pulse train, going from
$1.8U_p$, where $U_p$ is the ponderomotive energy, to vanishingly
small values. In the former case, this leads to strong harmonics
throughout, whereas the latter case causes enhancements only in
low-energy regions of the spectra.

In Ref. \cite{atthhg2004}, however, the modifications in the spectra
have been attributed to the selection of specific electron
trajectories, which are present in case it is released by a
high-frequency wave, and, subsequently, propagates in a strong
infra-red field. In comparison to the case in which the electron is
released only by the infrared field, there is a splitting in the
electron orbits, into the so-called "downhill" and "uphill"
trajectories, with respect to the effective potential barrier
$V_{\mathrm{eff}}=V(r)-\mathbf{r}.\mathbf{E}(t^{\prime})$ at the
electron start time $t^{\prime}$, where $\mathbf{E}(t^{\prime})$ and
$V(r)$ denote the electric field and the atomic binding potential,
respectively. The former and the latter case relate to the situation
for which the electron velocity, at $t'$, has the opposite or the
same direction of the field, respectively. If the time delays are
appropriately chosen, one may enhance ionization, and thus the HHG
spectra, for a particular set of orbits, or for none.

 Probably, such discrepancies are rooted on the fact
that the attosecond pulse train has been modeled in different ways.
While in \cite{atthhg2004} it has been approximated by a
monochromatic wave, in \cite{FSVL2006} we considered a Dirac-Delta
comb. In this context, one should keep in mind that a realistic
attosecond-pulse train possesses a broad frequency range. Therefore,
a monochromatic wave is as much as an approximation as a Dirac-Delta
comb. The main difference lies in the number of harmonics composing
the train: the former should work better if this number is small
(such as in \cite{atthhg2004}), and the latter if it is large.

In this proceeding, we present a detailed discussion of the two
limits which one may adopt, when modeling the attosecond pulse
train. We put a particular emphasis on the consequences of taking
the attosecond pulse train to be a monochromatic wave, within the
Strong-Field Approximation, for high-order harmonic generation. In
order to make an assessment of the similarities and differences from
the Dirac Delta comb employed in our previous work \cite{FSVL2006},
we will also briefly recall such results.

This manuscript is organized as follows: In Sec. \ref{theory}, we
present the explicit expressions for the transition amplitudes, if
the attosecond pulse train is taken to be a monochromatic wave (Sec.
\ref{mono}) or a Dirac Delta comb (Sec. \ref{deltcomb}). In
particular, we discuss how the saddle-point equations are modified
for each case, and relate their solutions to the classical orbits of
an electron in a field. Subsequently, in Sec. \ref{res}, we compute
such orbits explicitly, and employ them to calculate high-harmonic
spectra. Finally, in Sec. \ref{conclusions} we conclude the paper,
relating both limits to the results in \cite{atthhg2004}.

\section{Transition amplitudes}

\label{theory} The general expression for the HHG transition
amplitude, in the strong-field approximation \cite {footnsfa}, is
given by
\begin{eqnarray}
M_{\Omega } &\hspace{-0.1cm}=\hspace*{-0.1cm}&i\int_{-\infty }^{\infty
}\hspace*{-0.5cm}dt\int_{-\infty }^{t}~\hspace*{-0.5cm}dt^{\prime }\int
d^{3}kd_{z}^{\ast }(\mathbf{k}+\mathbf{A}(t))d_{z}(\mathbf{k}+\mathbf{A}%
(t^{\prime }))  \nonumber \\
&&E(t^{\prime })\exp [iS(t,t^{\prime },\Omega ,\mathbf{k})],  \label{amplhhg}
\end{eqnarray}
with the action
\begin{equation}
S(t,t^{\prime },\Omega ,\mathbf{k})=-\frac{1}{2}\int_{t^{\prime }}^{t}[%
\mathbf{k}+\mathbf{A}(\tau )]^{2}d\tau -I_{p}(t-t^{\prime })+\Omega t,
\label{action}
\end{equation}
where $\mathbf{A}(t),$ $\Omega ,\mathbf{k}$ and $I_{p}$ denote the vector
potential, the harmonic frequency, the intermediate momentum of the freed
electron and the atomic ionization potential, respectively \cite{hhgsfa}.
Eq. (\ref{amplhhg}) describes a physical process in which an electron is
freed at a time $t^{\prime }$, propagates in the continuum with momentum $%
\mathbf{k}$ from $t^{\prime }$ to $t$, and, at this time, recombines with
its parent ion, generating a harmonic of frequency $\Omega $. The prefactors
\begin{equation}
d(\mathbf{k}+\mathbf{A}(t))=\left\langle \psi _{0}\right| \mathbf{r}.\mathbf{%
\epsilon }_{x}\left| \mathbf{k}+\mathbf{A}(t)\right\rangle  \label{ff1}
\end{equation}
and
\begin{equation}
d^{\ast }(\mathbf{k}+\mathbf{A}(t^{\prime }))=\left\langle \mathbf{k}+%
\mathbf{A}(t^{\prime })\right| \mathbf{r}.\mathbf{\epsilon }_{x}\left| \psi
_{0}\right\rangle ,  \label{ff2}
\end{equation}
where $\left| \psi _{0}\right\rangle $ and $E(t^{\prime })=-dA(t^{\prime
})/dt^{\prime }$ denote the electronic bound state and the electric field at
the time the electron is released, respectively, contain all the influence
of the atomic binding potential, which is implicit in the bound-state wave
functions. In general, the vector potential is given by $\mathbf{A}(t)=%
\mathbf{A}_{l}(t)+\mathbf{A}_{h}(t),$ and the total electric field
by $\mathbf{E}(t)=\mathbf{E}_{l}(t)+\mathbf{E}_{h}(t)$, where the
indices $l$ and $h$ refer to the laser field and the attosecond
pulses, respectively. In this paper, we work within the so-called
``broad gaussian limit'', which implies that the electron is bound,
and returns to, a bound state localized at the origin of the
coordinate system \cite{hhgsfa}. This yields constant prefactors
(\ref {ff1}) and (\ref{ff2}).

For low enough frequencies and high enough laser intensities, Eq. (\ref
{amplhhg}) can be solved to a good approximation by the steepest descent
method. Thus, we must determine $\mathbf{k}$, $t^{\prime }$ and $t$ so that $%
S(t,t^{\prime },\Omega ,\mathbf{k})$ is stationary, i.e., so that
its partial derivatives with respect to these parameters vanish.
Apart from considerably simplifying the computations involved, this
approximation has the advantage of providing a clear space-time
picture for the physical process in question, and allowing a
detailed assessment of quantum-interference effects. Furthermore, as
it will be shown below, the saddle-point equations can be related to
those describing the classical orbits of an electron
\cite{orbitshhg}.

If the attosecond pulses are absent, the saddle-point equations read
\begin{equation}
\left[ \mathbf{k}+\mathbf{A}_{l}(t^{\prime })\right] ^{2}=-2I_{p}
\label{saddle1}
\end{equation}

\begin{equation}
2(\Omega -I_{p})=\left[ \mathbf{k}+\mathbf{A}_{l}(t)\right] ^{2},
\label{saddle2}
\end{equation}
and
\begin{equation}
\int_{t^{\prime }}^{t}d\tau \left[ \mathbf{k}+\mathbf{A}_{l}(\tau )\right]
=0.  \label{saddle3}
\end{equation}
Eq. (\ref{saddle1}) and (\ref{saddle2}) give the energy conservation at the
start and recombination times, respectively, while Eq. (\ref{saddle3})
yields the intermediate electron momentum fulfilling the condition for the
electron to return. The first equation expresses tunneling ionization at $%
t^{\prime },$ and has no real solution. Physically, this means that
tunneling has no classical counterpart. In the limit of
$I_{p}\rightarrow 0,$ this equation describes a classical electron
being released with vanishing drift velocity. Eq. (\ref{saddle2})
illustrates a process in which the kinetic energy of the electron
upon return is converted into a photon of frequency $\Omega $.\ Near
the cutoff, whose energy position corresponds to the maximal value
this quantity may take, such a process has no classical counterpart,
and the transition probability is expected to decay exponentially.

We will now include the attosecond-pulse train. The strong,
low-frequency field will be approximated by a linearly polarized
monochromatic wave of amplitude $E_{0}$, i.e.,
\begin{equation}
\mathbf{E}_{\mathrm{l}}(t)=E_{0}\sin \left( \omega t\right) \mathbf{\epsilon
}_{x},
\end{equation}
while the attosecond-pulse train will be taken as
\begin{equation}
\mathbf{E}_{\mathrm{h}}(t)=\frac{E_{\mathrm{h}}}{{\sigma(t)}}\sum_{q=2k_{0}+1}^{2k_{1}+1}\sin %
\left[ q\omega (t-t_{d})\right] \mathbf{\epsilon }_{x},
\label{attopulse1}
\end{equation}
of amplitude $E_{\mathrm{h}}$. Eq. (\ref{attopulse1}) is a sum over
several odd high-order harmonics, which are phase-locked and exhibit
a time delay $t_{d}$ with respect to the background field. The
indices $2k_{0}+1$ and $2k_{1}+1$ yield the minimal and the maximal
harmonic order, respectively. The function $\sigma(t)$ denotes the
train temporal envelope, which will be taken as $\sigma=const.$
Physically, this corresponds to an infinitely long attosecond-pulse
train.

 In our model, we assume that the attosecond pulses will mainly
influence the electron ejection in the continuum but not its
subsequent propagation. This latter step will be governed by the
low-frequency field. For this reason, unless stated otherwise, we
will use the approximation $E(t^{\prime })\simeq
E_{\mathrm{h}}(t^{\prime })$ in the prefactor of Eq.
(\ref{amplhhg}), but not in the action. This implies that the vector
potential is approximated by $\mathbf{A}(t)\simeq
\mathbf{A}_{l}(t).$

\subsection{The monochromatic limit}

\label{mono}

For a finite number of harmonics, this yields highly oscillating prefactors,
which are no longer slowly varying. Hence, they must be incorporated in the
action. The HHG transition amplitude then reads
\begin{equation}
M_{\mathrm{h}}=-\frac{E_{\mathrm{h}}}{2i\sigma}\sum_{q=2k_{0}+1}^{2k_{1}+1}M_{%
\mathrm{h}}^{(q)},  \label{matto}
\end{equation}

\bigskip with
\begin{eqnarray}
M_{\mathrm{h}}^{(q)}\hspace{-0.2cm} &=&\hspace{-0.2cm}\int_{-\infty
}^{\infty }\hspace{-0.3cm}dt\int_{-\infty }^{t}\hspace*{-0.3cm}dt^{\prime
}\int \hspace{-0.1cm}d^{3}kd^{\ast }(\mathbf{k}+\mathbf{A}_{l}(t))d(\mathbf{k%
}+\mathbf{A}_{l}(t^{\prime }))  \nonumber \\
&&\hspace*{-0.3cm}(e^{iq\omega (t^{\prime }-t_{d})}-e^{-iq\omega (t^{\prime
}-t_{d})})\exp [iS(t,t^{\prime },\Omega ,\mathbf{k})].
\end{eqnarray}
In Eq. (\ref{matto}), we will define a modified action,
\begin{equation}
\tilde{S}=S\pm q\omega (t^{\prime }-t_{d}).
\end{equation}
This leads to changes in the saddle-point equation (\ref{saddle1}), which
now reads
\begin{equation}
\left[ \mathbf{k}+\mathbf{A}_{l}(t^{\prime })\right] ^{2}=2(\mp q\omega
-I_{p}).  \label{sadd1mod}
\end{equation}
Thereby, the solution in $-q\omega $ does not make physically sense,
and thus will not be taken into account, whereas that in $+q\omega $
describes an ionization process in which an electron is ejected by
the absorption of a high-frequency photon. Hence, the
attosecond-pulse train, in which a superposition of such processes
occurs, provides an alternative pathway for the electron to reach
the continuum. This is in agreement with the discussions in
\cite{atthhg2004}. We will consider here the limit for which the
attosecond pulse train is described by a monochromatic wave, so that
the sum in (\ref{matto}) is dropped.

If Eq. (\ref{sadd1mod}) is combined with (\ref{saddle2}) and (\ref{saddle3}%
), one obtains the expressions
\begin{equation}
\omega t^{\prime }=\epsilon _{1}\arccos \alpha  \label{tstart1}
\end{equation}
and
\begin{eqnarray}
\epsilon _{2}\sqrt{1-\alpha ^{2}}-\sin \omega t &=&  \label{tret1} \\
&=&\omega (\epsilon _{1}\arccos \alpha -t)(\cos \omega t-\gamma _{2}),
\nonumber
\end{eqnarray}
with $\alpha =\cos \omega t+\gamma _{1}-\gamma _{2},$ and $\epsilon _{i}=\pm
1$ $(i=1,2),$ for the\ electron ionization and return \ times, respectively.
Hence, the problem of finding the pairs of times $(t,t^{\prime })$ has been
reduced to solving the transcendental equation (\ref{tret1}). In the
above-stated equations, one may identify the parameters
\begin{equation}
\text{ }\gamma _{1}=\pm \sqrt{\frac{\Omega -I_{p}}{2U_{p}}}\text{ and }%
\gamma _{2}=\pm \sqrt{\frac{q\omega -I_{p}}{2U_{p}}}.  \label{gammas}
\end{equation}
In particular $\gamma _{2}$ is very important for determining the initial
conditions for the electron when it is being ejected in the continuum.

One should note that there exist sign ambiguities in the above-stated
equations. Furthermore, additional complications may be caused by the fact
that $\arccos \alpha =$ $\arccos (2n\pi -\alpha ).$ We will discuss such
problems explicitly in Sec. \ref{orbs}, in which they are overcome by means
of physical arguments.

\subsection{The Dirac-Delta limit}

\label{deltcomb}

In the limit $k_{1}-k_{0}\longrightarrow \infty $, which is the main
assumption in our former work \cite{FSVL2006}, Eq. (\ref{attopulse1}) reads
\begin{equation}
\mathbf{E}_{\mathrm{h}}(t)=\frac{E_{h}\pi}{\sigma}
\sum_{n=0}^{\infty }(-1)^{n}\delta (\omega (t-t_{d})-n\pi
)\mathbf{\epsilon }_{x}. \label{attopulse2}
\end{equation}
 This yields the transition amplitude
\begin{eqnarray}
M_{\mathrm{h}}^{(D)} &=&\frac{i\pi E_{\mathrm{h}}}{\sigma }%
\sum_{n=0}^{\infty }(-1)^{n}\int_{-\infty }^{+\infty }\hspace*{-0.5cm}dt\int
d^{3}k\exp \left[ iS(t,t_{n}^{\prime },\Omega ,\mathbf{k})\right]  \nonumber
\\
&&d_{z}^{\ast }(\mathbf{k}+\mathbf{A}_{l}(t))d_{z}(\mathbf{k}+\mathbf{A}%
_{l}(t_{n}^{\prime })).  \label{hhgatto}
\end{eqnarray}
In this case, the ionization time is being fixed at $t_n^{\prime
}=t_{d}+n\pi /\omega $. The above-stated equation means, physically,
that the electron is no longer reaching the continuum through
tunneling ionization, but is being released by the attosecond
pulses. However, in contrast to the approximation in the previous
section, all information about the velocity with which the electron
is leaving is lost.
Indeed, the electron is reaching the continuum with any of the energies $%
N\omega -I_{p}$, since all harmonics composing the train are equivalent.

Under these assumptions, the transition amplitude is further simplified so
that only the integrals in the intermediate electron momentum $\mathbf{k}$
and the return time $t$ must be solved. In this case, the saddle-point
equations (\ref{saddle2}) and (\ref{saddle3}) can be combined in order to
obtain the transcendental equation
\begin{eqnarray}
&&\sin \omega t-(-1)^{n}\sin \omega t_{d}  \nonumber \\
&=&\left[ \omega (t-t_{d})-n\pi \right] \left( \cos \omega t-\gamma
_{1}\right) ,  \label{sadddelt}
\end{eqnarray}
where $\gamma _{1}$ is defined in (\ref{gammas}). Similarly to the
monochromatic-limit case, one may identify ambiguities in Eq. (\ref{sadddelt}%
). It turned out, however, that they are fewer and far easier to overcome in
the limit discussed in this Section.

\section{Results}

\label{res}

\subsection{Start and return times}

\label{orbs}

We will now assess the influence of the additional attosecond-pulse
train on the electron orbits. As a starting point, we will
approximate it by a high-frequency monochromatic wave. In general,
we will refer to a pair of orbits employing
the indices $(i,j)$, which increase with the electron excursion times $%
\Delta t= t-t^{\prime}$ in the continuum.

We will initially concentrate on the shortest pair of orbits, i.e.,
$(1,2)$, which provide the most prominent contributions to the
high-harmonic spectra. In Fig. 1, we display the real parts of the
ionization and return times for such orbits, obtained from the
saddle point equations (\ref{sadd1mod}), (\ref {saddle2}) and
(\ref{saddle3}), as compared to the situation for which the
attosecond pulses are absent (i.e., for $q=0$ in (\ref{sadd1mod})).
In all cases, start and return times occur in pairs, which coalesce
at well-defined energies. Such energies correspond to the maximal
kinetic energy with which a classical electron may return to its
parent ion. For higher energies, the transition amplitudes are
exponentially decaying. Hence, potentially, such energies lead to
cutoffs in the spectra. For a monochromatic field in the absence of
the attosecond-pulse train, the cutoff occurs at $%
3.17U_{p}+I_{p}.$

In the figure, we can identify two main distinct regimes. If the frequency $%
\Omega _{q}=q\omega $ is lower than the ionization potential
$I_{p}$, qualitatively, there is the same behavior for the start
times $t^{\prime }$ or return times $t.$ Indeed, the only difference
is the cutoff energy, which is slightly lower. An inspection of Eqs.
(\ref{tstart1}) \ and (\ref{tret1}) supports this interpretation: if
$q\omega -I_{p}<0$, the parameter $\gamma _{2}$ is purely imaginary,
so that the solutions of (\ref{tret1}) occur in conjugate pairs. The
element of such a pair which leads to $\mathrm{Im}[t^{\prime }]<0$
is discarded as unphysical, whereas the other element is directly
associated with the tunneling process. In this context, the quantity $\mathrm{Im}%
[t^{\prime }]$ provides information about how probable the tunneling
process is, as it will be discussed below.

\begin{figure}[tbp]
\begin{center}
\includegraphics[width=8cm]{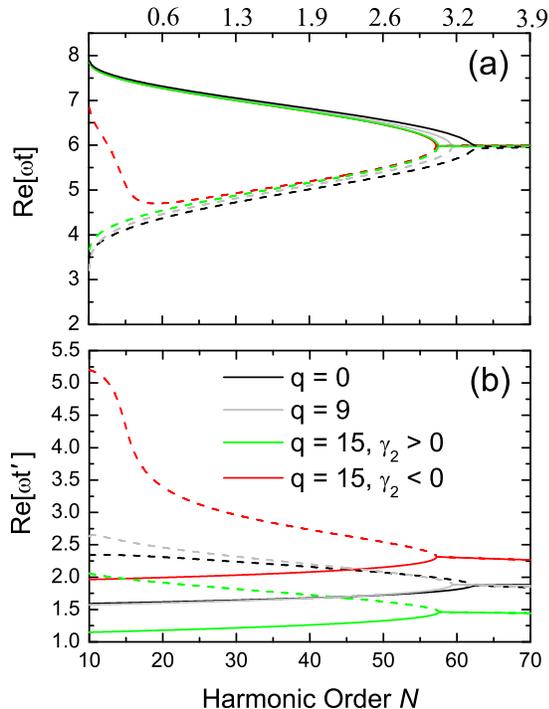}
\end{center}
\caption{(Color Online) Real parts of return and start times (panel
(a) and (b), respectively), as functions of the high-harmonic order,
for the two shortest
orbits of an electron subject to a monochromatic field of intensity $%
I=4\times 10^{14}\mathrm{W/cm^{2}}$, and frequency $\protect\omega
=0.057$ a.u., superposed to a high-frequency, monochromatic wave of
$I_{h}=I/10$, approximated by a monochromatic wave of frequency
$\Omega_q=q\protect\omega$. For comparison, we also provide such
times for the case in which the attosecond train is absent ($q=0$ in
Eq.(\ref{sadd1mod})). The atomic system is taken to be argon
($I_{p}=0.58$ a.u.) and we only show the harmonic orders for which
$\Omega >I_{p}$. The dashed and the solid lines in the figure refer
to the orbit $1$ and $2$, respectively. The numbers on the upper
edge of the figure denote the kinetic energy of the electron upon
return, in units of the ponderomotive energy $U_p$.}
\end{figure}
If on the other hand, $q\omega -I_{p}>0$ (for instance, for $q=15)$,
the attosecond pulses have given enough energy for the electron to
overcome the potential barrier. In this case, it is reaching the
continuum with velocity $v=\pm \sqrt{2(q\omega -I_{p})}.$ In
comparison to a situation for which the electron is being released
with vanishing velocity, which is the classical limit of Eq.
(\ref{saddle1}), there is a splitting in the ionization times
$t^{\prime }$ for each set of orbits. This is precisely the effect
reported in Ref. \cite{atthhg2004}, and can be readily understood
from Eq. (\ref{tret1}), in which now the parameter $\gamma _{2}$ is
real. Hence, there exist now two solutions of (\ref{tret1}) which
make physically sense. The solution with $\gamma _{2}<0$ yields the
downhill trajectories in \cite{atthhg2004}, whereas that with
$\gamma _{2}>0$ corresponds to the uphill trajectories.

Interestingly, however, for both types of trajectories, the electron
returns almost at the same time $t$. In fact, there are only
noticeable differences near the ionization threshold $\Omega
=10\omega $, which, at most, will influence the low-plateau region.
Hence,
the electron excursion times in the continuum are considerably shorter for $%
\gamma _{2}<0$. Obviously, the larger the frequency $\Omega
_{q}=q\omega ,$ the more pronounced the splitting is. In comparison
with \cite{atthhg2004}, however, there is a much larger overlap of
the start times for both types of trajectories near the ionization
threshold.

From a more technical viewpoint, it is worth mentioning that, in the
computation of the orbits $(1,2)$, we have taken $\epsilon _{1}=\epsilon
_{2}=+1$ and $\gamma _{1}<0$ in general. As an exception, for $q=15$, $%
\gamma _{2}<0$, and $i=1$, for harmonic frequencies $\Omega /\omega \leq
24.4 $ one must employ an analytical continuation in Eq. (\ref{tret1}), with
$\epsilon _{2}=-1,$ $\epsilon _{1}=+1$ and $2\pi -\arccos \alpha $ instead
of $\arccos \alpha ,$ in order to guarantee continuous solutions.

\begin{figure}[tbp]
\begin{center}
\includegraphics[width=8cm]{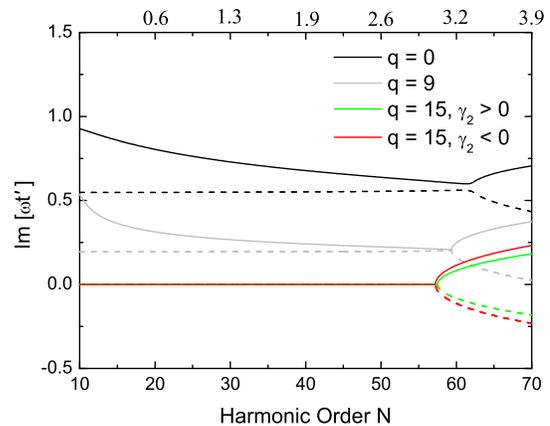}
\end{center}
\caption{(Color online) Imaginary parts of electron start times for
the two shortest orbits of an electron subject to the same field and
atomic parameters as in the previous figure. In the figure, we only
show the harmonic orders for which $\Omega
>I_{p}$. The dashed and the solid lines in the figure refer to the orbit $1$
and $2$, respectively. The numbers on the upper edge of the figure
denote the kinetic energy of the electron upon return, in units of
the ponderomotive energy $U_p$.}
\end{figure}

The above-stated physical interpretation is confirmed by Fig. 2, in which
the imaginary parts of the start times are depicted. Such imaginary parts
are in a sense a measurement of a quantum-mechanical process having a
classical counterpart. Indeed, the smaller $|\mathrm{Im}[t^{\prime }]|$ is,
the larger the probability that the process in question takes place will be.
In the figure, we observe that, for the monochromatic and $q=9$ cases, $|%
\mathrm{Im}[t^{\prime }]|\neq 0.$ This is expected, since tunneling
ionization has no classical counterpart. Furthermore,
$|\mathrm{Im}[t^{\prime }]|$ decreases if the attosecond pulses are
present, due to an effectively narrower barrier. In case the
electron is able to overcome the atomic binding potential (e.g., for
$q=15$), for energies lower than the cutoff, $|\mathrm{Im}[t^{\prime
}]|$ is vanishingly small, since now a classical counterpart does
exist. Clearly, beyond the cutoff energy, the imaginary parts of
such times increase.
\begin{figure}[tbp]
\noindent\includegraphics[width=9.4cm]{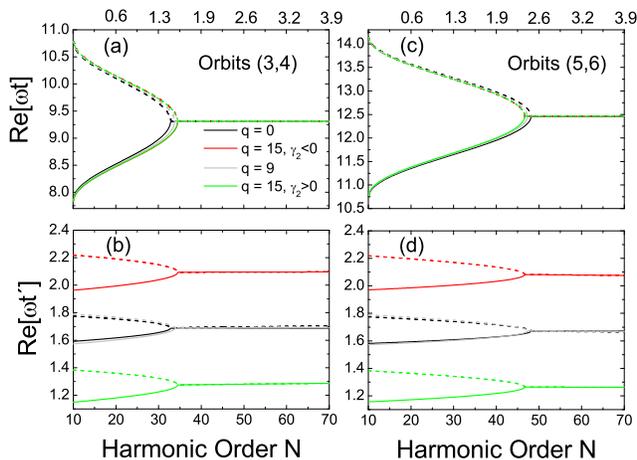}
\caption{(Color online) Real parts of start and return times for the
pairs of orbits $(3,4)$ [panels (a) and (b)], and $(5,6)$ [panels
(c) and (d)], as functions of the high-harmonic order, for the same
atomic and field parameters as in the
previous figures. The orbits $(3,4)$ have been obtained by taking $\protect%
\epsilon _{1}=\protect\epsilon _{2}=+1$ and $\protect\gamma _{1}>0$ in Eq. (%
\ref{tret1}), while $(5,6)$ have been found employing $\protect\epsilon _{1}=%
\protect\epsilon _{2}=+1$ and $\protect\gamma _{1}<0$. The numbers
on the upper edge of the figure denote the kinetic energy of the
electron upon return, in units of the ponderomotive energy $U_p$.}
\end{figure}

The splitting in the electron start times is not exclusive of
$(1,2)$, but, in fact, occurs for all pairs of orbits.  As concrete
examples, in Fig. 3 we illustrate the real parts of the start and
return times for the orbits $(3,4)$ and $(5,6).$ Such
orbits correspond to electron excursion times $\Delta t\sim 1.5T$ and $%
\Delta t\sim 2T,$ respectively, where $T=2\pi /\omega $ denotes a
cycle of the driving field. Clearly, the start times $t^{\prime }$
split for $q\omega -I_{p}>0,$ due to the electron non-vanishing
velocity upon ejection. Once more, each pair of start times leads to
a shorter and a longer set of orbits, which correspond to $\gamma
_{2}<0$ or $\gamma _{2}>0$, respectively. The corresponding electron
start times do not overlap in any energy region. This is due to the
fact that they are much more localized than for the orbits $(1,2).$

For comparison, in Fig. 3, we present the solutions of Eq.
(\ref{sadddelt}), obtained when the attosecond-pulses are
approximated by a sum of Dirac-Delta functions, for several time
delays. The start and return times, as well as the cutoff energies,
are different from those in the previous figures. Indeed, we no
longer observe a splitting in the sets of ionization and
recombination times, as compared to the situation for which the
attosecond pulses are absent. This is in perfect agreement with the
fact that now there is no longer a well-defined velocity with which
the electron is being ejected. In contrast, however, there is now a
single ionization time per half cycle, determined by Eq. (\ref
{attopulse2}). The return times, however, still occur in pairs that
coalesce at the cutoffs. For each pair of orbits, the cutoff
energies, as well as the excursion times, considerably changes with
the time delay $t_d$.
\begin{figure}[tbp]
\hspace*{-1cm} \includegraphics[width=9.5cm]{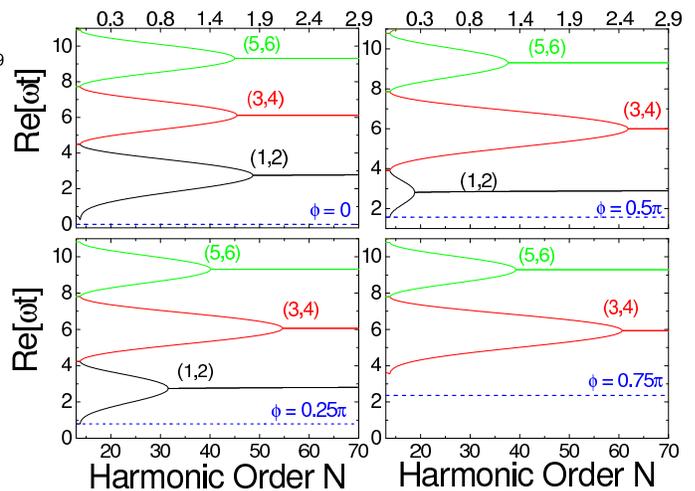}
\caption{(Color online) Real part of the recombination times as a function
of the high-harmonic order, for neon ($I_{p}=0.79$ a.u.) interacting with a
monochromatic field of intensity $I=5\times 10^{14}\mathrm{W/cm^{2}}$, and
frequency $\protect\omega =0.057$ a.u. and an attosecond-pulse train of
intensity $I_{\mathrm{h}}=10^{13}\mathrm{W/cm^{2}}$, composed of an infinite
number of harmonics (Eq. (\ref{attopulse2})). Parts (a), (b), (c) and (d)
correspond to delays $\protect\phi =0$, $\protect\phi =0.25\protect\pi $, $%
\protect\phi =0.5\protect\pi $ and $\protect\phi =0.75\protect\pi $
between the attosecond-pulse train and the low-frequency driving
wave, respectively. In the figure, we only show the harmonic orders
for which $\Omega >I_{p}$, and the pairs of orbits are indicated by
the natural numbers $(i,j)$, which increase according to the
electron excursion time in the continuum. The ionization times are
indicated by the blue dashed lines, and the numbers on the upper
edge of the figure give the approximate kinetic energy of the
electron upon return, in units of the ponderomotive energy $U_{p}$.}
\label{hhgfig2}
\end{figure}

\subsection{Harmonic spectra}

We will now analyze the spectra computed under the assumptions in
Sec. \ref{mono}, i.e., that the attosecond-pulse train can be
approximated by a high-frequency, monochromatic wave. For
comparison, we will also present spectra computed considering that
the APT is a sequence of Dirac-Delta functions. In this latter case,
however, we will keep the discussion as brief as possible, as a
detailed analysis is already presented in \cite{FSVL2006}. In both
cases, we have employed a uniform saddle-point approximation in
order to compute the transition amplitudes. This approximation is
discussed in detail in \cite{atiuni}.

If one concentrates on the physical picture of downhill and uphill
trajectories, a very important question is the influence of each
type of orbit in the high-harmonic spectra, and how such
trajectories can be selected. In \cite{atthhg2004},
such a selection has been performed by employing an appropriate time-delay $%
t_{d}$ between the strong, infra-red field and the attosecond-pulse
train. In our framework, however, since we are approximating the
train by a monochromatic wave, all information about this parameter
is lost. However, we can mimic the effect achieved in
\cite{atthhg2004} by leaving out, or including, the corresponding
orbits in our computations.

In Fig. 5.(a), we present the spectra computed for a train with the
frequency $\Omega _{15}=15\omega $ employing all orbits up to
$(5,6)$. In this case, $q\omega -I_{p}>0$, so that the electron is
being ejected with non-vanishing drift velocity and the spectra
contains contributions from both the uphill and the downhill
trajectories. We have restricted the electron start times to the
first half cycle of the laser field, so that there are no sharp
harmonics in the figure. One is able to view, however, the
structures caused by the interference between the different pairs of
trajectories very clearly. Prominent features are also a cutoff near
harmonic order $N=57$. This is in agreement with Fig. 1, which shows
that the pairs of orbits (1,2) coalesce around this energy, and also
with Fig. 2, for which there is a sharp increase in
$|\mathrm{Im}[t^{\prime }]|$ at this harmonic order. Furthermore,
there is a very pronounced hump in the low-plateau region.
\begin{figure}[tbp]
\noindent \includegraphics[width=9cm]{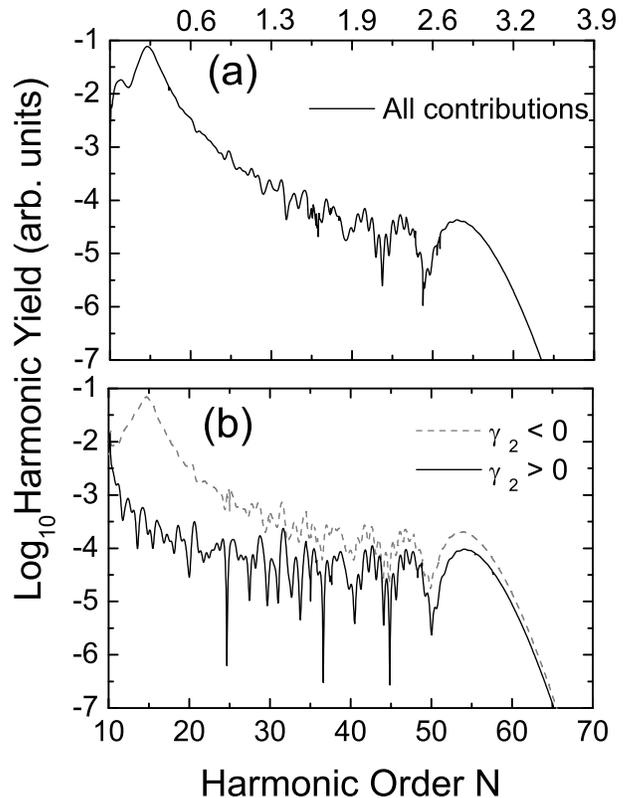}
\caption{Harmonic spectra computed with the orbits $(1,2)$, $(3,4)$ and $%
(5,6)$, for an attosecond-pulse train of frequency $\Omega _{15}=15\protect%
\omega $ superposed to a strong, near-infrared field of frequency $\protect%
\omega =0.057$ a.u. The field and atomic parameters are the same as
in Fig. 1. Panel (a) exhibits the contributions from all sets of
orbits, while panel (b) displays the individual contributions from
the orbits with $\protect\gamma _{2}>0$ \emph{or} $\protect\gamma
_{2}<0$ in Eq. (\ref{tret1}). The numbers on the upper edge of the
figure give the kinetic energy of the electron upon return, in units
of the ponderomotive energy $U_p$.}
\end{figure}

More detail is provided in Fig. 5.(b), in which the individual
contributions from the downhill ($\gamma _{2}<0$) and the uphill
($\gamma _{2}>0$) trajectories are displayed. In general, the former
contributions are at least half an order of magnitude larger than
the latter. This is expected, since, for $\gamma _{2}<0,$ the
electron excursion times in the continuum are shorter for all sets
of orbits. This means that there is less spreading for the
electronic wave packet, and consequently a larger overlap, upon
return, between such a wave packet and the atomic bound state with
which it recombines. Obviously, this leads to a larger harmonic
yield. Furthermore, in both cases there is no difference in the
cutoff energy. This is due to the fact that all trajectories split
symmetrically, with respect to the purely monochromatic case.

The above-stated features can be understood in detail analyzing the
contributions from each pair of orbits. Such contributions are
displayed in Fig. 6. For both cases, the main features in the
spectra, such as their overall intensities, the plateau shapes and
cutoff energies, are determined by the orbits $(1,2)$. In fact, the
contributions from such orbits are at least one order of magnitude
larger than those from $(3,4)$ or $(5,6)$. This is in particular
true for the case $\gamma _{2}<0$, due to the shorter excursion
times. Apart from that, they coalesce at a much larger energy than
the other sets. This means that, classically, the maximal energy for
which the electron returns to its parent ion along $(1,2)$, and
consequently the cutoff, is larger than for the longer orbits. In
the figure, we also notice that the hump in the low-plateau region,
for $\gamma _{2}<0,$ can be traced back to the fact that the
electron excursion time in the continuum, for the set $(1,2)$ is
extremely short in this case (see, e.g., Fig. 1). The other pairs of
orbits mainly contribute to the substructure in the spectra. Another
noteworthy feature is that, regardless of the sign of $\gamma_2$,
the cutoff energies are the same for all sets of orbits, in
agreement with the previous discussion.

\begin{figure}[tbp]
\noindent \includegraphics[width=9cm]{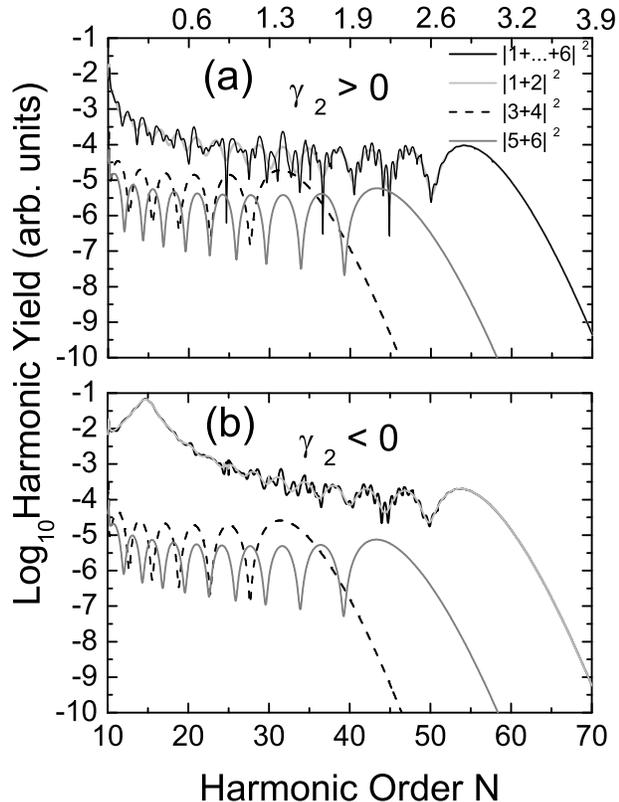}
\caption{Individual contributions from the shortest three pairs of
orbits to the high-harmonic spectra, for the same field and atomic
parameters as in the previous figure. Panel (a) \ and (b) depicts
the contributions from $\protect\gamma _{2}>0$ or $\protect\gamma
_{2}<0$ in Eq. (\ref{tret1}), respectively. For comparison the
overall contributions are displayed as the solid black lines in the
figure. The numbers on the upper edge of the figure give the kinetic
energy of the electron upon return, in units of the ponderomotive
energy $U_p$.}
\end{figure}

Another scenario, presented in Fig. 7 for comparison, is observed if
the attosecond-pulse train is taken to be a Dirac-Delta comb. In
this case, the dominant set of orbits, and, consequently, the shape
of the spectra, are highly dependent on the time delay $t_d$. For
instance, if the time delay is vanishing [Fig. 7.(a)], the spectrum
is dominated by the orbits $(1,2)$. This is due to the fact that
they coalesce at the energy $I_p+1.8U_p$, which is at a relatively
high harmonic order. Furthermore, since the electron excursion times
are very short, the harmonics are quite prominent. For $\phi=\omega
t_d=0.25\pi$ [Fig. 7.(b)], there is now a double plateau, with two
consecutive intensity drops at $I_p+0.93U_p$ and $I_p+2.11U_p$.
These are the energies for which the orbits $(1,2)$ and $(3,4)$
coalesce, respectively. Specifically in the upper half of the
plateau, there was a drop in one order of magnitude, as compared to
Fig. 7.(a). This is due to the larger excursion times, and,
consequently, wave-packet spreading, for $(3,4)$. If the delay
increases further [Fig. 7.(c)], the contributions from $(1,2)$ only
cause a shoulder in the spectrum, due to the fact that they now
coalesce at $I_p+0.26U_p$, and the plateau, extending until
$I_p+2.5U_p$, is determined by the orbits $(3,4)$. The latter orbits
dominate the spectra even more for $\phi=\omega t_d=0.75\pi$ [Fig.
7.(d)], since the cutoff energy for $(1,2)$ is now vanishingly
small. This effect is very similar to that observed in
\cite{atthhg2004}, even though it occurs for completely different
delays.

\begin{figure}[tbp]
\hspace*{-1cm}
\includegraphics[width=10cm]{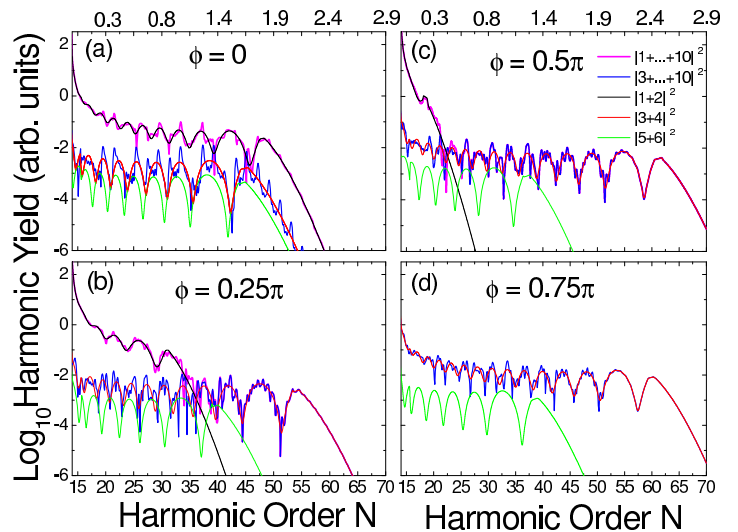}
\caption{ (Color online) Contributions of isolated pairs of orbits
to the high-harmonic spectra for the same atomic and external-field
parameters in Fig.~4, and delays $\phi=0$, $\phi=0.25\pi$,
$\phi=0.5\pi$ and $\phi=0.75\pi$ between the attosecond-pulse train
and the low-frequency driving wave
(Figs.~\ref{hhgfig3}(a),~\ref{hhgfig3}.(b),~\ref{hhgfig3}.(c)
and~\ref{hhgfig3}.(d), respectively). In the figure, we only show
the harmonic orders for which $\Omega>I_p$. The contributions from
specific pairs of orbits are depicted in the same colors as those
employed in Fig. 4, for depicting each corresponding pair. For
comparison, the spectra computed with the ten shortest orbits and
with orbit 3 to 10 are given as the thick lines in the figure.  The
numbers on the upper edges of the figure give the approximate
kinetic energy of the electron upon return, in units of the
ponderomotive energy $U_p$.} \label{hhgfig3}
\end{figure}
\section{Conclusions}

\label{conclusions}

 The outcome of this work supports the results in
\cite{atthhg2004} in several ways. In general, it also shows that,
by changing the initial conditions with which an electron is ejected
in the continuum, one may manipulate several features in the
spectra, such as the high-harmonic intensities, or cutoff.
Furthermore, by considering the electron to be released in the
continuum by a monochromatic wave, we have obtained a splitting of
the electron start times very similar to that reported in
\cite{atthhg2004}, within the context of the Strong-Field
Approximation, in the so-called uphill and downhill trajectories.

 We have also observed enhancements in the low-plateau region, and higher
 overall intensities for the contributions from the downhill
trajectories.  This indicates a good perspective for harmonic
control, by selecting either the downhill or uphill trajectories
with, for instance, an adequate delay between the infrared field and
the attosecond-pulse train.

We have also obtained very detailed information about the
contributions of specific pairs of orbits to the spectra. In the
monochromatic limit, the shortest pair of trajectories, denoted
$(1,2)$ is completely dominant, determining the shape of the
plateau, the cutoff energy and the overall intensity in the spectra.
This is due to the fact that the maximal classical energy for this
pair, and thus the cutoff, is always at a higher harmonic order than
for the longer pairs of trajectories.

In contrast, in case one considers a sum of Dirac-Delta functions,
the dominance of this pair will depend very much on the time delay
between the infra-red field and the attosecond pulses. This is due
to the fact that the maximal energy an electron returning along
these orbits is not always at a
higher harmonic frequency than for the longer pairs. For instance, if for $%
\omega t_{d}=0,$ the orbits (1,2) coalesce at the end of the plateau, and
thus overshadow the remaining contributions, for  $\omega t_{d}=0.25\pi $
and $\omega t_{d}=0.5\pi ,$ this energy lies at the low-energy part of the
plateau, or near the ionization potential, respectively. Hence, the orbits
(3,4) will also play an important role in the spectra.

A feature in common for both the monochromatic and the Dirac-Delta
limits is that the excursion time $\Delta t=t-t^{\prime}$ of the
electron in the continuum  is a very important parameter, and
considerably influences the shape of the plateau and harmonic
intensities. This is not surprising, since, now, for all sets of
orbits, the electron is reaching the continuum with a roughly equal,
and large, probability. Hence, the differences in the yield will be
mainly determined by the overlap between the returning electronic
wave packet and the bound state with which it recombines. The
shorter the orbit along which the electron is returning is, the
larger this overlap will be. In contrast, if the electron is
released by tunneling ionization, the probability with which it is
ejected will depend much more critically on the instantaneous
potential barrier, and also influence the yield.

Moreover, it is remarkable that both asymptotic limits exhibit
features in common with those obtained in \cite{atthhg2004} within a
more realistic model for the attosecond pulses. The monochromatic
limit, for instance, allows one to identify the downhill and uphill
trajectories, and to analyze their consequences in the spectra.
Approximating the attosecond pulse train by a sum of Dirac Delta
functions, on the other hand, sheds light on several features
reported in  \cite{atthhg2004}, such as a double plateau, or changes
of more than one order of magnitude in the harmonic spectra.
However, only in Delta-Dirac case there is a decrease in the energy
of the enhanced harmonics, with increasing time delays $t_d$.

There exist, however, aspects of \cite{atthhg2004} that we have not
addressed. We have not, for instance, investigated the differences
in the high-harmonic resolution, due to a particular trajectory
choice. Indeed, in order to do so, it would have been necessary to
extend the start times to several cycles of the laser field.
Instead, we have restricted them only to the first half cycle, in
order to have a closer look at quantum interference effects.
Furthermore, the uniform approximation employed in this work
requires that we deal with pairs of trajectories (for details see
\cite {atiuni}), and such effects are expected to be related to
single trajectories in a pair \cite{saclay96}. These issues were not
the main objective of this work, and are open for further studies.

\acknowledgements{We are grateful to M. Lewenstein for useful
discussions. This work has been financed in part by the U.K. EPSRC
(Advanced Fellowship, grant No. EP/D07309X/1).}


\begin{thebibliography}{99}
\bibitem{footnsfa}  The SFA consists in neglecting the atomic potential in
the propagation of the electron in the continuum, the laser field when the
electron is bound or rescatters, and the excited states of the atom in
question.
\end{thebibliography}
\end{document}